\documentclass[12pt,a4paper]{article}
\usepackage{graphicx}
\usepackage{xcolor}
\usepackage{times}
\title{\bf Global performance and capabilities of the instruments on the 3.6-m Devasthal Optical Telescope}

\author{Amitesh Omar\thanks{email: aomar@aries.res.in (AO)}, Bheemireddy Krishna Reddy, Tripurari Kumar,\\ Jayshreekar Pant 
\vspace{0.5cm}\\
\normalsize Aryabhatta Research Institute of observational sciences, Manora Peak, Nainital 263001 India
}
\date{\mbox{}}
\begin{document}
\maketitle
\setcounter{page}{1001}
\pagestyle{plain}
    \makeatletter
    \renewcommand*{\pagenumbering}[1]{%
       \gdef\thepage{\csname @#1\endcsname\c@page}%
    }
    \makeatother
\pagenumbering{arabic}

%
%
\def\bull{\vrule height .9ex width .8ex depth -.1ex}
\makeatletter
\def\ps@plain{\let\@mkboth\gobbletwo
\def\@oddhead{}\def\@oddfoot{\hfil\scriptsize\bull\quad
"2nd Belgo-Indian Network for Astronomy \& astrophysics (BINA) workshop'', held in Brussels (Belgium), 9-12 October 2018 \quad\bull}%
\def\@evenhead{}\let\@evenfoot\@oddfoot}
\makeatother
%
%
\def\beginrefer{\section*{References}%
\begin{quotation}\mbox{}\par}
\def\refer#1\par{{\setlength{\parindent}{-\leftmargin}\indent#1\par}}
\def\endrefer{\end{quotation}}
%
%
{\noindent\small{\bf Abstract:} 
The recently commissioned 3.6-m Devasthal optical telescope has been used for various tests and science observations using three main instruments, namely, a charge-coupled device camera, a near-infrared camera, and an optical imager-cum-spectrograph. The published results from these instruments assert that the performance of the telescope at the Devasthal site is at par with the expectations. These back-end instruments open up vast opportunities for high-sensitivity observations of the celestial sky with the telescope. This paper provides a summary of the existing back-end instruments and attempts to highlight the importance of the Devasthal optical telescope in synergy with other telescopes operating at different wavelengths.
}
\vspace{0.5cm}\\
{\noindent\small{\bf Keywords:} Devasthal Optical Telescope -- astronomical instrumentation -- spectrograph -- imaging -- high speed imaging}
%
%
\section{Introduction}

The Devasthal Optical Telescope (DOT) with its 3.6-m primary mirror is the largest optical telescope in India (Sagar 2017). The DOT became operational in 2016 (Kumar et al. 2018a) and is operated and managed by the Aryabhatta Research Institute of observational sciences (ARIES), Nainital, an autonomous research institute under the Department of Science \& Technology (DST), Government of India. The telescope is located at Devasthal ($79.69^{0}$ E, $29.36^{0}$ N, 2426 m above the mean sea level) along with two other optical telescopes, namely, the 1.3-m telescope installed in 2010 (Sagar et al. 2011) and the upcoming 4-m international liquid mirror telescope (Surdej et al. 2018). The DOT was built by AMOS, Belgium (Ninane et al. 2012a; Flebus et al. 2012). As per statement released at the time of the remote technical activation of the DOT in March 2016 ({\it http://pib.nic.in/newsite/PrintRelease.aspx?relid=138446}), the aim of the DOT is to provide a world-class, ground-based observing facility for spectroscopic exploration and imaging  of the Milky Way and celestial objects at optical wavelengths as well as to carry out follow-up studies of the objects detected at radio wavebands by the Giant Metrewave Radio Telescope (GMRT, Swarup et al. 1991; Gupta et al. 2017), Pune and at UV and X-ray wavebands by the AstroSat, the first dedicated Indian multi-wavelength astronomical satellite (Agarwal 2017). 

Several publications highlight the design and technical details as well as the results of tests and characterization of the DOT (Flebus et al. 2012; Ninane et al. 2012a, 2012b; Ninane et al. 2016; Kumar et al. 2016; Kumar et al. 2018a), and the scientific importance of the DOT (Sagar 2000, 2017; Omar et al. 2017). In this paper, we  provide a summary of the most resent results obtained from the observations carried out using the DOT. We also briefly discuss the scientific potential of the existing instruments in synergy with other telescopes. 

\section{The salient features of the DOT observing facility} 

In this section, technical features and advantages of the DOT facility are briefly described. The geographical advantages of the DOT are also highlighted. Most of the information given below is taken from the publications mentioned before.

\subsection{The telescope}

Some key characteristics of the DOT facility relevant for the planning of science observations are described below. \\

\noindent {\bf Optical system:} The DOT with the Ritchey-Chr\'etien optics has the Cassegrain focal-plane at f/9 with a plate-scale of $6.366$~arcsec~mm$^{-1}$. It has three observing ports, namely, one main port and two side-ports. The side-ports provide $10'$ diameter clear field-of-view (FoV) via the light reflected from the side-port fold-mirror. The main port provides $30'$ diameter FoV, however, an annular part between $10'-30'$ may get vignetted due to the pick-off-mirror (POM) in the guider arm, if employed for the closed-loop tracking of the telescope. The extent of the vignetting in the $30'$ FoV depends on the location of the POM within the annular region between $31'-35'$. It may be noted that the POM can also be moved anywhere in the main FoV between $0'-30'$ and hence its position should be checked before starting the science observations. In case, no auto-guiding is required, POM should be retracted in a parking position beyond the $35'$ field. The 80\% encircled energy values from the optics are measured to be less than $0.4''$. The ratio of the stray light reaching the focal plane and the sky background is expected to be about one part in 1000 in the optical bands. The primary mirror has been re-aluminized twice at Devasthal after its first aluminization in 2015, using a magnetron-based coating system (Krishna Reddy et al. 2016).  \\

\noindent {\bf Mechanical and opto-mechanical systems:} The DOT is mounted on an altitude over azimuth mount. While tracking an object in the sky, the focal plane is continuously de-rotated to compensate the field rotation caused by the azimuth motion of the telescope. The primary mirror is actively controlled using 69 actuators to maintain its correct shape and position at all the elevations. The secondary mirror attached to a hexapod system is also actively controlled in position and orientation to correct the optical aberrations arising from temperature variations, mechanical flexure of the tube and active optics system-related corrections. During the observations,  a calibrated look-up table is used to make corrections to the active optics system. If desired, the parameters may also be determined on-the-fly using the wave-front sensor (WFS), which is optically co-aligned with the auto-guider. The auto-guider uses light between 550-750 nm while the WFS uses the remaining light (400-550 nm and 750-950 nm), taken from the POM in the annular region around the main FoV. The pointing accuracies with the telescope pointing models are measured to be nearly $1.3''$ rms. As the best seeing-limited images have a Full-Width at Half-Maximum (FWHM) of $0.4''$ (Omar et al. 2017, Kumar et al. 2018a), the closed-loop tracking system should normally be used during long exposures. The closed-loop tracking using the guider has been shown to be excellent, i.e. within $0.1''$ rms up to 1 hour of exposure.  \\

\noindent {\bf Operation:} The DOT is operated via a   Telescope Control Software (TCS). The software allows to monitor vital parameters of the telescope sub-systems and the positions of its axes. The target position can be fed via the TCS either directly on the TCS graphical user interface (normal user mode) or via a dedicated TCP/IP link using a pre-defined set of commands (engineering mode). The focal plane can be rotated arbitrarily within the allowed range to facilitate slit-based spectroscopic observations. The dome slit of the telescope enclosure tracks the telescope's position (Gopinathan et al. 2016). 

\subsection{Enclosure}

The cylindrical enclosure hosting the DOT is made of steel so as to minimize the thermal load on the telescope. The inside of the dome has insulation and the outer surface is reflective in order to minimize heating in the daytime. The telescope floor is at 11 m height from the local ground level and has 12 ventilation fans (Kumar et al. 2018b). The operating fans suck outside air from the dome slit and other openings (e.g., windows of non-operating fans) and flushes the air out to the surrounding. These fans are operated in the evening hours before the start of observations. Usually, only a few fans are operated depending upon the wind direction. These fans help in obtaining fast equalization of the temperature inside the telescope enclosure with the outside temperature. This thermal equilibrium improves the dome seeing. The windows of the fans are also usually left open during the observations to maintain a passive air flow. The technical room at the ground floor hosting heat-generating telescope sub-systems (e.g., chiller, compressor etc.) has a separate air exhaust system, which prevents the hot air to rise to the telescope floor. These arrangements are crucial in getting sub-arcsec seeing from the DOT.

\begin{table}
\caption{Technical summary of the existing back-end instruments on DOT.
\label{table1}} 
\small
\begin{center} 
\begin{tabular}{| l | l l l |}
\hline 
{\bf Parameter} & {\bf CCD Imager$^{1}$} & {\bf TIRCAM-2$^{2}$} & {\bf ADFOSC} \\
\hline
Wavelength coverage & 0.35--1$\mu$m & 1--3.7$\mu$m & 0.35--1$\mu$m\\
 &  &  & \\ 
Imaging device & Blue-enhanced Si-CCD & InSb FPA  & $a.$ Deep-depletion Si-CCD\\
& & & $b.$ EM frame-transfer CCD \\
Native pixel-scale & $0.095"$ & $0.169"$ & $0.20"~(a)$ \\
 &  &  & $0.107"~(b)$ \\ 
Field-of-view & $6.5'\times6.5'$ & $1.44'\times1.44'$ & $ 13.6'\times13.6'~(a)$ \\ 
 &  &  &$1.8'\times1.8'~(b)$ \\ 
Broad-band filters & UBVRI, $ugriz$ & J, H, K & $ugriz$, BG-39$^{*}$, RG-610$^{*}$ \\
 &  &  & \\  
Narrow-band filters & - & Br-$\gamma$, K-cont, PAH, & H$\beta^{\dagger}$ (486 - 538 nm)$|$ [$\Delta\lambda \sim $ \\ 
 &  & nbL & H$\alpha^{\dagger}$ (656 - 738 nm)$|$ $10~nm$] \\ 
Dispersing elements & - & - & 0.1 - 0.23 nm/pxl grism\\ 
 &  &  & 6 nm/pxl prism\\ 
Slits & - & - & $0.4'' - 1.6''\times8'$ \\ 
 &  &  & \\ 
$<10$ ms frame-rate$^{@}$ & - & Yes & Yes (with EM CCD)  \\ 
 &  &  & \\ 
Faintest detection (reported) & B=$24\pm0.2$ mag & J=$19\pm0.1$ mag & $i=25\pm0.3$ mag \\
 &  &  &$g=19$~mag (0.23 nm/pxl) \\  
Minimal FWHM (reported) & $1.2''$ &$0.45''$ in K-band &$1.1''$ in $z$-band \\
\hline 
\end{tabular} 
\end{center} 
Note: Parameters taken from $^{1}$Pandey et al. (2018) and  $^{2}$Baug et al. (2018). $^{*}$BG-39 and RG-610 are used as order-sorting filters with passbands of 350-600 nm and 600-1100 nm respectively. $\dagger$Several H$\alpha$ and H$\beta$ filters each having bandwidth $\sim$10 nm are available covering the wavelength-range provided in the table. $^{@}$Frame-rate possible with windowing and on-chip binning.
\end{table} 

\subsection{DOT's position in an international context}

The global importance of the optical telescopes located in India are well described in Sagar (2017). Globally, about twenty moderate-size (3-m to 4.3-m) optical observing facilities are presently operated across the world. The DOT falls in the middle of this aperture range. Out of these twenty telescopes, seven are in the mainland USA, five are in Chile, three are in Hawaii (Mauna Kea), three in western Europe (Canary islands, Spain) and one in eastern Australia and one in India.   The DOT therefore fills a large longitudinal gap for the 3-4 m class observing facilities between eastern Australia and western Europe. It is also within the latitude of 30$^{0}$ North, thereby providing a full coverage of the northern hemisphere sky and a large part of the southern sky with a good visibility to the Galactic center and the southern Galactic plane.  As some science cases require continuous or time-critical observations of faint variable objects or transient sources detectable only by 3-4 meter class observing facilities, the geographical location of Devasthal makes the DOT an internationally important observing facility.

\section{Existing back-end instruments}

\subsection{Optical CCD imager}

The optical charge-coupled device (CCD) imager, sensitive at the visible wavelengths, is an imaging instrument for the main port using the native f/9 focal plane of the DOT. It is equipped with several broad-band filters and a 4Kx4K CCD camera procured from Semiconductor Technology Associates, USA. The details of this instrument and the first results are given in Pandey et al. (2018). Some technical specifications of the CCD imager are given in Table~1.  

\subsection{TIRCAM-2}

The TIFR Near Infrared Imaging Camera - II (TIRCAM-2) is an imaging instrument provided by the Tata Institute of Fundamental Research (TIFR), Mumbai. The details of TIRCAM-2 and first results are given in Baug et al. (2018). Some technical specifications of the TIRCAM-2 are given in Table~1.  The observations taken with TIRCAM-2 mounted at the main port indicate that sources down to 19.0, 18.8, and 18.0 mag can be detected in the J, H, and K bands respectively with nearly 10\% photometric precision. The best seeing obtained was at nearly $0.45''$ in the K-band.

\subsection{ADFOSC}

\begin{figure}[h]
\centering
\includegraphics[width=13cm]{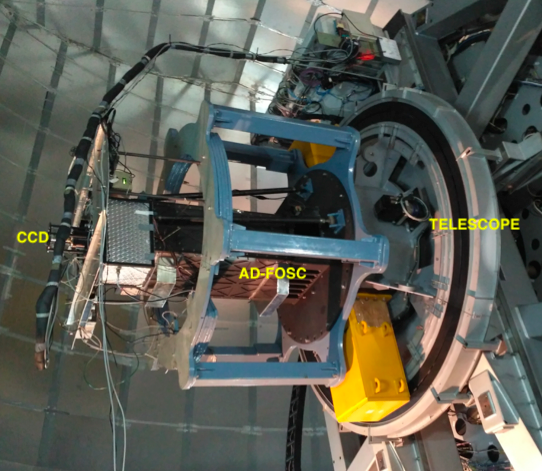}
\caption{The ADFOSC back-end instrument mounted on the DOT.}  
\label{fig_1}
\end{figure}

\begin{figure}
\centering
\includegraphics[width=12cm]{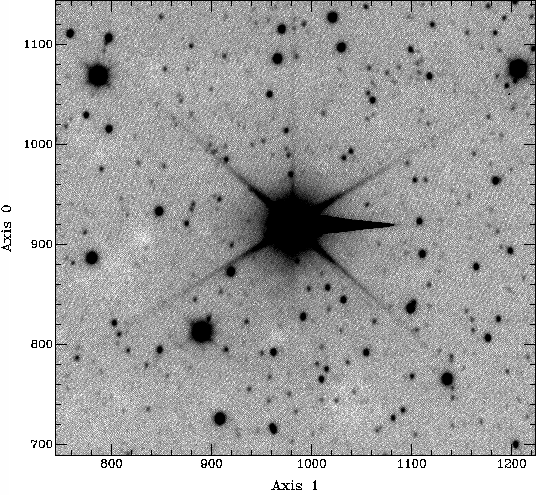}
\caption{The diffraction pattern around stars caused by the secondary mirror enclosure and the spiders supporting it. The thick horizontal conical trail emanating from the star to the right, unrelated to the diffraction pattern, is due to CCD charge-bleeding. The spikes rotate in the field with respect to the stars with time as the tracking progresses and therefore may cause spurious variabilities in nearby stars. It may also be noted that no significant ghost and excessive scattering around bright stars and  reflections elsewhere in the field (i.e. stray light) are visible. This image is taken with the ADFOSC on DOT. The image size is nearly $3'\times3'$.} 
\label{fig_4}
\end{figure}

\begin{figure}
\centering
\includegraphics[width=14cm]{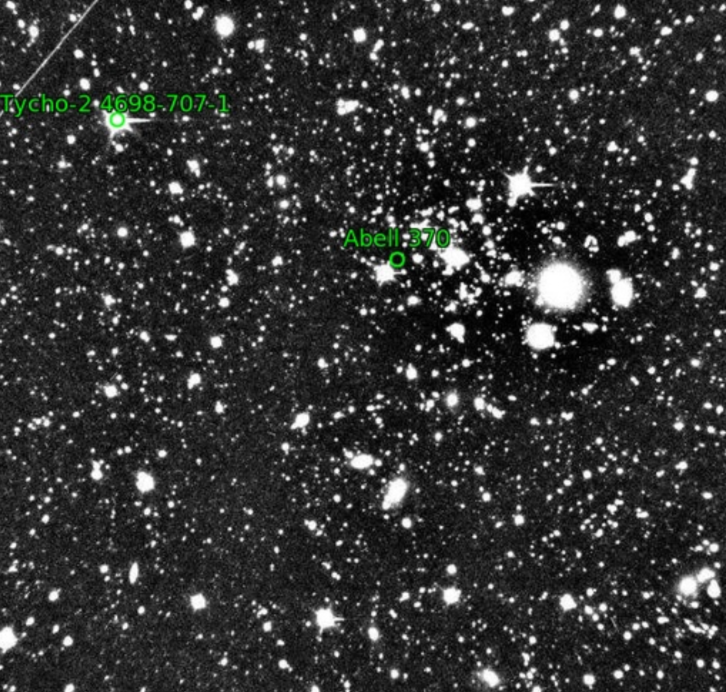}
\caption{A deep $i$-band image of the Abell 370 cluster field made using ADFOSC in a total of 55 minutes of integration time with the co-adding of 11  frames. The detection sensitivity is close to i=25 mag$_{AB}$ with a photometric precision of 0.3 mag. The uniformity of the sky background across the field is at a level of 0.1\% of rms. The image size is nearly $12'\times12'$.}  
\label{fig_2}
\end{figure}

\begin{figure}
\centering
\includegraphics[width=10cm]{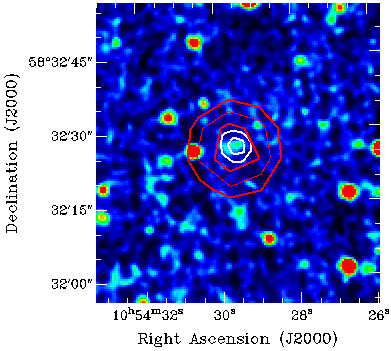}
\caption{The detection of a high-redshift ($z=4.8\pm2$) radio galaxy candidate in $i$-band image with $24.3\pm0.2$ mag$_{AB}$) using ADFOSC in about 1 hour of integration time (see Omar et al. 2019 for details). The red contours mark the GMRT 150 MHz detection and white contours mark the Very Large Array 1400 MHz detection of the radio galaxy.}  
\label{fig_3}
\end{figure}

The ARIES Devasthal - Faint Object Spectrograph \& Camera (ADFOSC), a focal-reducing instrument for the main port, is capable of performing both imaging and spectroscopy of celestial objects in the optical bands. ADFOSC is based on the design concept of the highly versatile FOSC instruments developed by the European Southern Observatory (ESO) and the University of Copenhagen (Buzzoni et al. 1984, Anderson et al. 1995). The details of ADFOSC are provided in Omar et al. (2012, 2019a). ADFOSC is equipped with several broad-band filters, narrow-band filters, slits, grisms and prisms. It has an internal calibration unit for spectroscopic and flat-field calibrations. Some parameters of the instrument are given in Table~1.

A view of  ADFOSC  mounted on the DOT is shown in Figure 1. It can be configured with either the 4Kx4K CCD or the electron-multiplying (EM) frame-transfer CCD. The 4Kx4K CCD camera used in ADFOSC is developed at ARIES, Nainital in collaboration with the Herzberg Institute of Astrophysics, Canada. The 4Kx4K CCD (back-illuminated E2V sensor with a peak quantum efficiency 90\%) is cooled using a closed-cycle heat exchanger system to -120 $^{0}$C while the EM-CCD (front-illuminated Texas Instruments sensor with a peak quantum efficiency 65\%) procured from ANDOR (UK) is thermo-electrically cooled to nearly -20 $^{0}$C. The readout noise in the EM-CCD camera is $<1 e^{-}$ rms at high EM gain values while that for the 4Kx4K CCD camera is nearly $e^{-}$ rms. The frame-transfer EM-CCD camera with a readout rate of 13.5 MHz with a possibility to read a small sub-frame with binning is suitable for fast-imaging (up to a few ms) of variable sources while the 4Kx4K camera with slow readout rate ($<200$ kHz) is suitable for wide-field deep imaging and spectroscopy. Presently, several upgrades are in advanced stages of integration in ADFOSC. A global positioning system (GPS) based time-stamping of the CCD images with an accuracy better than 0.1 ms is being tested. This will help executing time-coordinated science observations with different observatories. The arrangements are also being made to monitor the sky and perform auto-guiding (in addition to DOT's internal auto-guider) using an auxiliary CCD camera imaging about $1.9'\times1.4'$ of the sky at the f/9 focal plane, beyond that used for the 4Kx4K CCD camera. The external auto-guiding will help in correcting apparent tracking errors due to ADFOSC mechanical flexure.

The optical quality in terms of light scattering of ADFOSC and DOT system is satisfactory. As can be seen in Figure 2, the diffraction patterns due to the secondary mirror enclosure and its support system are minimal and other effects (such as reflections, stray light, ghosts) around  bright stars are either absent or not substantial in ADFOSC mounted on the DOT. ADFOSC has made deep images down to $\sim$25 mag$_{AB}$ in $i$-band with excellent photometric accuracy (0.2 - 0.3 mag and flat-fielding 0.1\% of rms) on the Abell 370 galaxy cluster field (see Figure 3). This indicates that ADFOSC is capable of making deep images with good photometric precision. ADFOSC will also be used to assist observations with the integral field spectrograph, under assembly and integration at the Inter-University Center for Astronomy and Astrophysics, Pune (Chung et al. 2014, 2018).

\section{Synergistic science areas} 

The DOT is expected to be an important facility for carrying out a wide variety of science observations of  interest to the Indian and Belgian astronomy communities. It has a large potential for use in synergy with telescopes in other wavebands. Before the next-generation large-area ($\sim10^{4}$ deg$^{2}$) deep-sky surveys such as PanSTARRS, the Dark Energy Survey, the Large Synoptic Survey Telescope reaching limiting magnitudes in the range 25-27 mag$_{AB}$ become publicly available, the 4-m class telescopes will continue to be used for obtaining deep photometric images of the sky. The deep imaging observations of  transients such as supernovae, Gamma-ray bursts (GRB), and gravitational-wave sources and the fields not covered by the sky surveys will, however, always be required using the 4-m class telescopes including the DOT. The major observational requirement at present seems to be follow-up spectroscopic observations in both optical and NIR bands. Spectroscopic observations of different types of sources and deep photometric observations of transients are to remain at the forefront for the DOT. The narrow-band imaging capability available with ADFOSC is also likely to be used heavily as no other high-sensitivity large-area  narrow-band imaging surveys are available for the moment ({\it in the future, the J-PAS survey using the Javalambre Survey Telescope is expected to provide narrow-band imaging data across the entire visible wavelength domain}). 

Several science areas proposed with the optical CCD imager and TIRCAM-2 are already described in earlier publications (e.g., Ojha et al. 2018, Pandey et al. 2018). The science areas include star clusters, star-forming regions, transient objects, extra-solar planets, and distant galaxies. Most of the science objectives planned with the optical CCD imager can also be fulfilled with ADFOSC. ADFOSC will also be used to obtain spectroscopy of transient objects such as supernovae and GRB. Here, we highlight some science areas for the DOT, particularly using ADFOSC, in synergy with other telescopes.

\subsection{Synergy with GMRT and other radio telescopes}
The GMRT and other low-frequency radio telescopes such as LOFAR (Low-Frequency Array; van Haarlem et al. 2013) and the upcoming SKA (Square Kilometer Array; Diamond \& Gupta 2017) are likely to detect several faint radio sources which will require optical cross-matched detections as well as spectroscopic observations to obtain redshifts. The majority of these radio sources are expected to be intermediate ($z\sim$3) and high-redshift ($z\geq5$) radio galaxies including a large population of star-forming galaxies. These objects are expected to be optically faint and due to the $k$-correction, relatively brighter in the optical-red ($r, i, z$) and NIR bands. The deep-depletion, fringe-suppressed CCD used in ADFOSC has higher detection sensitivity in the red bands up to 1$\mu$m, compared to  conventional CCDs. With typical $i-K$ colors expected between 2 and 5 mag, a K=20 mag object will have an i-band magnitude of 22 - 25 mag. ADFOSC has the capability of obtaining deep images with a precision of about 0.2 mag for a 24 magnitude object in about 1 hour of integration.  Recently, Omar et al. (2019b) reported the DOT $i$-band detection of a high-redshift candidate radio galaxy  discovered from GMRT data (see Figure 4). Therefore, searches for such red objects in the $i$ and $z$ bands with ADFOSC are expected to be as fruitful as that in the NIR bands using TIRCAM-2. Some of the bright objects (nearly 22 mag in $g$-band) can also be followed-up in optical spectroscopy using ADFOSC to obtain redshifts and study the properties of galaxies.

\subsection{Synergy with AstroSat}
AstroSat, India's first multi-wavelength satellite operating at X-ray and UV bands is routinely used for observations of transients and variable sources (e.g., black-hole candidates, X-ray binaries, pulsar, AGN), and nearby galaxies (Singh \& Bhattacharya 2017). The coordinated and follow-up observations in the optical/NIR bands are expected to be fruitful in several cases where multi-wavelength studies are essential to understand the radiation emission mechanisms and accretion phenomena. The DOT presently equipped with its frame-transfer EM-CCD with spectroscopic capabilities (using ADFOSC) has potential for coordinated observations. Using the EM-CCD on the DOT, single optical pulses from the Crab pulsar have been successfully detected. Figure 5 shows the frames with the detection of the pulsed emission from the Crab pulsar using the DOT. 

\begin{figure}
\centering
\includegraphics[width=14cm]{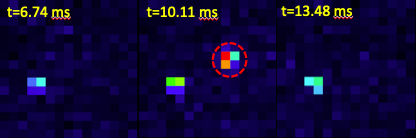}
\caption{The pulsed emission from the Crab pulsar (in red circle at t=10.11 ms) detected using the EM-CCD camera directly mounted on the DOT.}  
\label{fig_5}
\end{figure}

\subsection{Synergy with SDSS-derived products}

\begin{figure}
\centering
\includegraphics[width=14cm]{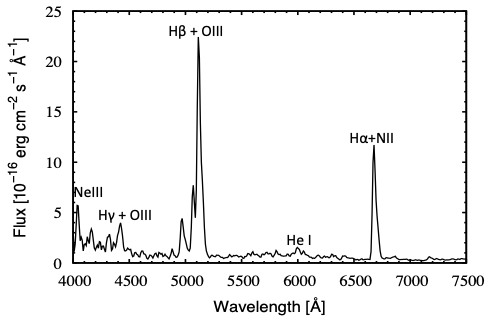}
\caption{The optical spectrum of a faint (g-band $\sim$19 mag) emission-line compact galaxy HS 2236+1344 recorded using ADFOSC with 600 seconds of integration time. }  
\label{fig_6}
\end{figure}

\begin{figure}
\centering
\includegraphics[width=12cm]{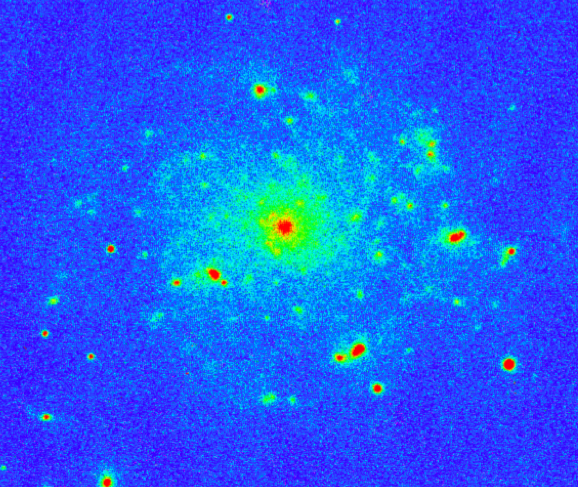}
\caption{The H$\alpha$ narrow-band (FWHM = 10~nm) image of galaxy NGC 3423 taken using ADFOSC mounted on the DOT. The image is not corrected for underlying continuum emission. Most of the discrete blobs showing intense emission are star-forming regions in the galaxy. The exposure time is 300 seconds.}  
\label{fig_7}
\end{figure}

The Sloan Digital Sky Survey (SDSS) has provided a great wealth of data and new sources which require follow-up spectroscopic and deeper photometric observations. Although the SDSS provides spectroscopic observations, these are often limited to brighter objects and a single point fibre-based spectrum.  The majority of the sources detected within the limiting sensitivity of the SDSS can be observed in spectroscopy using a 4-m class optical telescope. An area where the DOT has immense potential is the spatially-resolved chemical abundance mapping of nearby galaxies, particularly low-mass systems such as dwarf galaxies (e.g. see Paswan \& Omar 2019). To highlight the scientific potential of the DOT, an optical spectrum of a 19 mag dwarf galaxy taken using ADFOSC on the DOT is displayed in Figure 6. In addition, there are several SDSS-selected candidate quasar pairs and dual AGNs in galaxies which can be confirmed spectroscopically using the DOT (Chand, H., {\it private communication}). 

The AGN reverberation mapping to infer the black-hole mass and the geometry of the broad-line region using imaging in narrow-band filters (such as H$\alpha$ and H$\beta$) and also using spectroscopic monitoring is another area where DOT can be useful. A program on AGN reverberation mapping is currently being pursued using the ARIES telescopes (Chand et al. 2018) and several redshift-tuned narrow-band filters have been procured under this project (see Table 1). These filters can also be used for other science programs. Due to its moderate aperture, the DOT is suitable for narrow-band imaging of several objects including galaxies. One such example is the H$\alpha$-band image of a nearby galaxy NGC~3423 made using ADFOSC on the DOT shown in Figure 7. 

\section{Summary}

The DOT is presently equipped with three back-end instruments which are capable of providing high-sensitivity optical and NIR observations of the celestial sky. A summary of the current capabilities at the DOT and discussions on possible scientific synergies using facilities at other wavebands were presented. The preliminary results published  are asserting that the performance of the DOT is as expected from this 4-m class telescope at Devasthal.  

%
%
\section*{Acknowledgements}
We thank the anonymous referee and guest editor Patricia Lampens for a careful reading of the manuscript and useful suggestions. The authors thank colleagues too many to name individually here, for discussions on various topics related to the DOT, back-end instrumentation and prospective science topics which may be carried out using the DOT. This paper is written with a due acknowledgement of the efforts put in by various teams towards  instrumentation, operation and management of the DOT observing facility. AO thanks the organizers of the BINA workshop for the invitation to write this article. 
 
%
%
%

\footnotesize
\beginrefer

\refer Agrawal P. C. 2017, JApA, 38, 27 

\refer Andersen, J. et al. 1995, Msngr, 79, 12

\refer Baug T., Ojha D. K., Ghosh S. K. et al. 2018, JAI, 7, 1850003-1881

\refer Bheemireddy K., Gopinathan M., Pant J. et al. 2016, SPIE, 9906, 44

\refer Buzzoni, B. et al. 1984, Msngr, 38, 9 

\refer Chand H., Rakshit S., Jalan P. et al. 2018, BSRSL, 87, 291

\refer Chung H., Ramaprakash A. N., Omar A. et al. 2014, SPIE, 9147, 0V

\refer Chung H., Ramaprakash A. N., Khodade P. et al. 2018, SPIE, 10702, 7A

\refer Diamond, P., Gupta, Y. 2017, CSci, 113, 649

\refer Flebus C., Gabriel E., Lambotte S. et al. 2008, SPIE, 7012, 08

\refer Gopinathan M., Sahu S., Yadava S. et al. 2016, SPIE, 9913, 153  

\refer Gupta Y., Ajithkumar B., Kale H. S. et al. 2017, CSci, 113, 707

\refer Kumar B., Omar A., Gopinathan M. et al. 2018a, BSRSL, 87, 29

\refer Kumar T. S., Bastin C., Kumar B. 2016, SPIE, 9906, 3X

\refer Kumar T. S., Yadava S., Singh A. K. 2018b, in proceedings of 5th IEEE Uttar Pradesh Section International Conference on Electrical, Electronics and Computer Engineering (UPCON), 1

\refer Ninane N., Flebus C., Kumar B. 2012a, SPIE, 8444, 1V

\refer Ninane N., Bastic C., de Ville J. et al. 2012b, SPIE, 8444, 2U

\refer Ninane N., Bastin C., Flebus C. et al. 2016, SPIE, 9906, 4E

\refer Ojha D. K., Ghosh S. K., Sharma S. et al. 2018, BSRSL, 87, 58

\refer Omar A., Yadav R. K. S., Shukla V. et al. 2012, SPIE, 8446, 14

\refer Omar A., Kumar B., Gopinathan M. et al. 2017, CSci, 113, 682  

\refer Omar A., Kumar T. S., Reddy B. K. et al. 2019a, CSci, 116, 1472

\refer Omar A., Saxena A., Chand K. et al. 2019b, JApA, 40, A9

\refer Pandey S. B., Yadav R. K. S., Nanjappa N. et al. 2018, BSRSL, 87, 42

\refer Paswan A., Omar A. 2019, MNRAS, 482, 3803

\refer Sagar R. 2000, CSci, 78, 1076

\refer Sagar R., Omar A., Kumar B. et al. 2011, CSci, 101, 1020

\refer Sagar R., Kumar B., Omar A. et al. 2012, SPIE, 8444, 1T

\refer Sagar R. 2017, in proceedings of National Academy of Sciences, India - Sect. A, 87, 1

\refer Singh K. P., Bhattacharya D. 2017, CSci, 113, 602

\refer Surdej J., Hickson P., Harmanno B. et al. 2018, BSRSL, 87, 68

\refer Swarup G., Ananthakrishnan S., Kapahi V. K. et al. 1991, CSci, 60, 95

\refer van Haarlem M. P. et al. 2013, A\&A, 556, A2

\refer Kuhn J. R. et al. 2001, PASP, 113, 1486

\refer Toma R. et al. 2016, MNRAS, 463, 1099 

\endrefer   

\section*{Additional Note}

\subsection*{Limitations due to Alt-Az mount}

One serious limitation on observations using the Alt-Az mount comes from rotation of the point spread function (PSF) in the image-plane as an object is tracked or observed over a time period. While the field of view is rotated using the instrument de-rotator, the spikes in the PSF (see Fig. 2) caused by the fixed obstructions in the optical path in the telescope is rotated on the CCD plane. As a result, the spikes can introduce spatio-temporally variable background sky near other stars in the field. If this effect is not properly taken care of via some sort of de-convolution or identifying and flagging of the affected time range, spurious variabilities in the objects can result. This effect is described in detail in several publications, e.g., Kuhn et al. 2001, Toma et al. 2016. Users need to be therefore cautious while using DOT to study faint-level variability in crowded fields or in fields near bright stars. Observing with a fixed de-rotator angle may help in some situations. 

\end{document}